\documentclass[epjCONF,columns]{svjour} %for 2 columns format
\usepackage{graphics}
\usepackage[varg]{txfonts} % Times fonts
\usepackage[latin1]{inputenc}
\session-title{HCP 2011}
\begin{document}
\title{W/Z properties (except mass) form ATLAS and CMS}
\author{J. Santaolalla (on behalf of CMS and ATLAS collaborations)}
\institute{Centro de Investigaciones Energ\'eticas MedioAmbientales y Tecnol\'ogicas, Avda. Complutense, 22 - 28040 Madrid}
\abstract{
The results on W and Z boson properties by both ATLAS and CMS (except mass) are presented in this document. The inclusive W and Z boson cross section production, the W charge asymmetry, the differential production as a function of the boson rapidity and transverse momentum, the W polarization and the sinus of the weak angle are shown in this document. The studies included are based on LHC collisions at $\sqrt{s}=7$ TeV, recorded during 2010 and 2011.
} %end of abstract
\maketitle
\section{Introduction}
\label{intro}
The measurement of electroweak boson properties in CMS~\cite{atlas} and ATLAS~\cite{cms} is of capital importance, especially in the first years of operation of the LHC. These measurements are useful to test the Standard Model (SM) at a new energy scale, to improve our knowledge on the proton composition. They are also sensitive to new physics. They are important background for many searches. As standard candles, they are useful to calibrate our detectors. Additionally, electroweak bosons are copiously produced in the collisions, being one of the most common processes, and their signal is considerably clean in the leptonic channels (especially in the Z boson case). 

\begin{figure}
% Use the relevant command for your figure-insertion program
% to insert the figure file.
% For example, with the option graphics use
\resizebox{0.75\columnwidth}{!}{
  \includegraphics{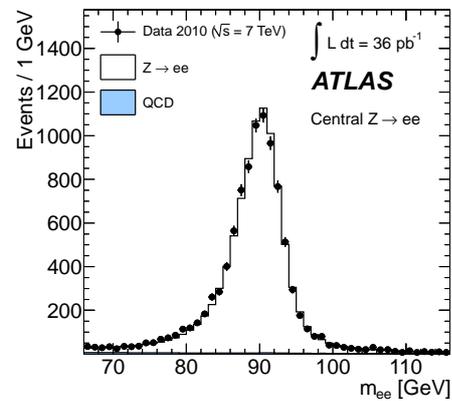} }
\caption{Z boson invariant mass distribution in the electron channel (ATLAS).}
\label{zATLAS}       % Give a unique label
\end{figure}

All the studies presented here share similar criteria for W and Z selection and reconstruction. Z bosons decay into two electrons or two muons approximately $10\%$ of the time. These events are characterized by two high $p_T$, isolated electrons or muons. The strategy in both experiments to select Z boson events is based on these characteristics, looking for energetic leptons (muons or electrons) not reconstructed within a jet. Quality criteria applied on the leptons are necessary to avoid bad reconstructions, bad $p_T$ measurements and to remove punch-through, cosmic rays and decays in flight. The invariant mass distribution for Z bosons decaying into two electrons after selection is shown in Fig.~\ref{zATLAS}, in the ATLAS case. The distributions in the muon channel and in the CMS case are similar. W bosons decay into one electron (muon) plus an electronic neutrino (muonic neutrino) approx. $10\%$ of the times per channel. In both cases we consider in the final state the existence of an energetic lepton (muon or electron) plus an imbalance of energy in the transverse plane (missing transverse energy, MET). The reconstruction of such a missing energy points generally to the existence of an undetected neutrino. The W boson signal is not as clean as the Z boson one due to the presence of the neutrino. Extra requirements are applied to remove backgrounds: quality criteria and electron $p_T$ (muon $p_T$) requirements (similar to the Z boson ones), Drell-Yan veto (removing events with two energetic leptons, to avoid Z bosons to be reconstructed as W bosons). In order to get a cleaner sample of W bosons, several studies require the transverse mass of the system lepton-neutrino, defined as:
\begin{equation}
  M_T = \sqrt{2p_T(\mu)\rm{MET}(1-\cos(\Delta\phi_{\mu \rm{MET}}))}
\end{equation}
to be higher than a certain value, generally 40-50 GeV (this latter requirement removes almost completely the multijet background, dominant background in the low $M_T$ region). In this equation, $p_T(\mu)$ is the muon transverse momentum, MET the missing transverse energy and $\Delta\phi_{\mu\rm{MET}}$ the angle between the muon and the MET.

%\begin{figure}
% Use the relevant command for your figure-insertion program
% to insert the figure file.
% For example, with the option graphics use
%\resizebox{0.75\columnwidth}{!}{
%  \includegraphics{incCMS.eps} }
%\caption{$M_T$ distribution of the neutrino-muon system after the W selection (CMS).}
%\label{incCMS}       % Give a unique label
%\end{figure}

\section{Inclusive electroweak boson production}
\label{sec:1}
\begin{figure}
% Use the relevant command for your figure-insertion program
% to insert the figure file.
% For example, with the option graphics use
\resizebox{0.75\columnwidth}{!}{
  \includegraphics{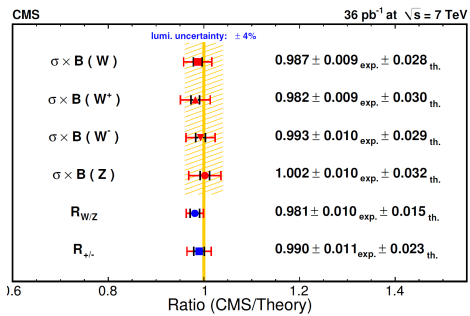} }
\caption{Ratio of results of CMS with respect to the theoretical prediction. All the results are in agreement with the SM prediction (CMS).}
\label{wzratioCMS}       % Give a unique label
\end{figure}

The electroweak boson production is studied in both CMS and ATLAS in the three leptonic channels. The W and Z boson inclusive cross section production through their decay into electrons are muons is the first electroweak measurement published by both experiments (~\cite{incCMS}, ~\cite{incATLAS}). It is measured by counting the number of events reconstructed as W or Z bosons for a certain luminosity and correcting for lepton reconstruction and identification efficiencies and detector acceptance. In the W boson case, the selection is optimized to get a reasonably clean sample consisting of more than $80\%$ of signal in the Jacobian peak region. The analysis slightly differs in CMS and ATLAS in this measurement. ATLAS performs a more restrictive selection (removing events with low $M_T$ or low MET). In CMS a fit to the MET variable is used to extract the signal normalization. The Z boson case is easier since it is almost background free (less than $1\%$). The results published are presented in the form of total and fiducial cross sections and ratios of cross sections. This latter measurement is a more stringent test of the SM since many systematic uncertainties are removed or reduced when computing the ratios (e.g. the dominant luminosity uncertainty). The total systematic uncertainty (excluded luminosity) is of the order of $2\%$ for both experiments. All the results are in agreement with the SM prediction. A summary of results published by CMS is shown in Fig.~\ref{wzratioCMS}.

\begin{figure}
% Use the relevant command for your figure-insertion program
% to insert the figure file.
% For example, with the option graphics use
\resizebox{0.75\columnwidth}{!}{
  \includegraphics{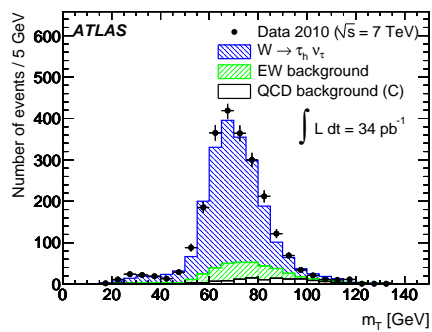} }
\caption{$M_T$ distribution for the decay of the W boson into $\tau$ and neutrino for data (dots), signal (blue), electroweak background (green) and QCD background (white) (ATLAS).}
\label{wtATLAS}       % Give a unique label
\end{figure}

\begin{figure}
% Use the relevant command for your figure-insertion program
% to insert the figure file.
% For example, with the option graphics use
\resizebox{0.75\columnwidth}{!}{
  \includegraphics{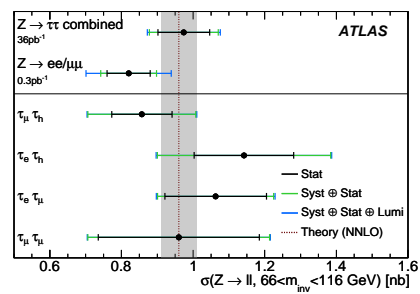} }
\caption{Z boson cross section production studied in the $\tau$ channel in the four final states: $\tau_{\mu}\tau_{\mu}$, $\tau_{\mu}\tau_{e}$, $\tau_{\mu}\tau_{had}$,  $\tau_{e}\tau_{had}$ and compared with the theoretical prediction (Gray band) and the cross section given in other leptonic decays (ATLAS).}
\label{ztATLAS}       % Give a unique label
\end{figure}

W and Z bosons are also studied in the tau decay mode. The W boson production cross section when decaying into $\tau$ and a $\nu_{\tau}$ neutrino is measured by the ATLAS collaboration~\cite{incWtATLAS} (CMS~\cite{incWtCMS} also published preliminary results). The properties of the tau jet (when the tau decays hadronically) are exploited to reconstruct these events. They select events with low multiplicity, high collimated jets or with non-zero flight distance using a boosted decision tree that includes all these properties. A clear W to tau signal is observed (more than $80\%$ signal). The main difficulty in this analysis is the electroweak background (W boson decays into electrons and muons and Z boson leptonic decays) that amounts to some $13\%$ and has a similar shape in the transverse mass as the signal (see Fig.~\ref{wtATLAS}). The cross section is measured with an approx. $23\%$ systematic uncertainty and is compatible with the SM prediction. The Z boson signature in the tau channel is cleaner. It is studied in both CMS~\cite{incZtCMS} and ATLAS~\cite{incZtATLAS} in four final states: $\tau_{\mu}\tau_{\mu}$, $\tau_{\mu}\tau_{e}$, $\tau_{\mu}\tau_{had}$,  $\tau_{e}\tau_{had}$. The electroweak background is partially removed by inverting the $M_T$ and MET requirements. The  $\tau_{l} \tau_{had}$ channel is the one with the biggest branching ratio, but we have to cope with important backgrounds, mainly QCD multijet events. The $\tau_e\tau_{\mu}$ is the cleanest since no electroweak process can produce both leptons in the same event apart from the rare diboson processes. The  $\tau_{\mu}\tau_{\mu}$ is highly affected by the $Z\rightarrow \mu\mu$ process. A boosted decision tree in 5 kinematic variables sensitive to the kinematical differences of muons from both processes is used by ATLAS. CMS uses a likelihood in 4 variables to separate signal and electroweak background. A competitive result is obtained after combining the results coming from the 4 final states (see Fig.~\ref{ztATLAS}). The total cross section is measured with approx. $10\%$ uncertainty by both experiments with a central value compatible with the SM prediction.

\subsection{Differential distributions}
\label{sec:2}
\begin{figure}
% Use the relevant command for your figure-insertion program
% to insert the figure file.
% For example, with the option graphics use
\resizebox{0.75\columnwidth}{!}{
  \includegraphics{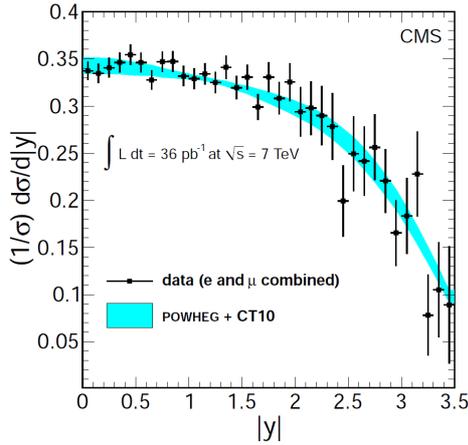} }
\caption{The normalized differential cross section for Z bosons as a function of the absolute value of rapidity, combining the muon and electron channels. The error bars correspond to the experimental statistical and systematic uncertainties.}
\label{zrapCMS}       % Give a unique label
\end{figure}

The differential distributions as a function of boson rapidity and momentum give valuable information, especially on the proton composition (PDFs). The Z boson production is studied in bins of the boson rapidity.  The procedure for the cross section measurement is similar to that of the total inclusive production. Results for both muons and electrons, and the combination of them are presented by both collaborations. In CMS~\cite{ZrapidityCMS} and ATLAS~\cite{incATLAS} they exploit the use of the forward calorimeters to reconstruct electrons at high rapidities. In this region the resolution is worse but provides information for the PDFs in a new ($Q^2$-$x$) region. The results are compatible with the expectations given by the reference MC in CMS and ATLAS. Systematic uncertainties range from $3\%$ in the central region to $7\%$ in the forward. Fig.~\ref{zrapCMS} shows the differential rapidity distribution for Z bosons in CMS.

\begin{figure}
% Use the relevant command for your figure-insertion program
% to insert the figure file.
% For example, with the option graphics use
\resizebox{0.75\columnwidth}{!}{
  \includegraphics{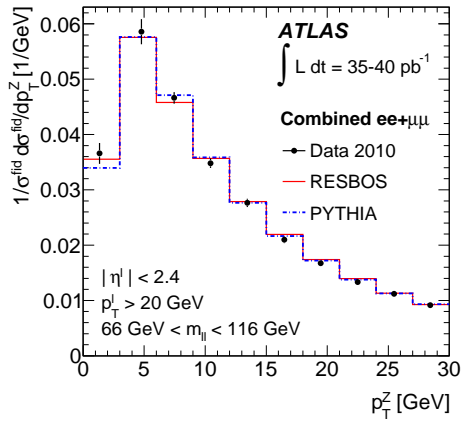} }
\caption{The combined normalized differential cross section as a function of $p_T$ (Z) for the range $p_T$(Z)$<$30 GeV compared to the predictions of RESBOS and PYTHIA. The error bars shown include statistical and systematic uncertainties (ATLAS).}
\label{zptATLAS}       % Give a unique label
\end{figure}

\begin{figure}
% Use the relevant command for your figure-insertion program
% to insert the figure file.
% For example, with the option graphics use
\resizebox{0.75\columnwidth}{!}{
  \includegraphics{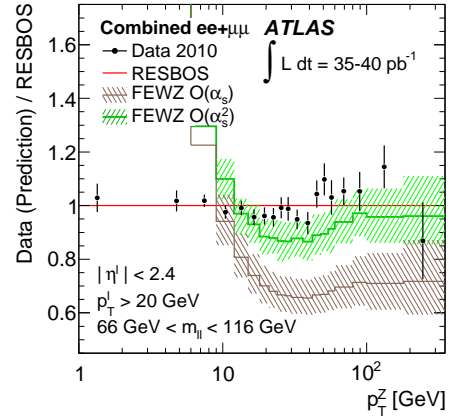} }
\caption{Ratios of the combined data and the FEWZ predictions at NLO and NNLO over the RESBOS prediction for the normalized differential cross section as a function of $p_T$(Z). The FEWZ predictions are shown with combined scale, $\alpha_s$, and PDF uncertainties. The data points are shown with combined statistical and systematic uncertainty (ATLAS). }
\label{zpthighATLAS}       % Give a unique label
\end{figure}

\begin{figure}
% Use the relevant command for your figure-insertion program
% to insert the figure file.
% For example, with the option graphics use
\resizebox{0.75\columnwidth}{!}{
  \includegraphics{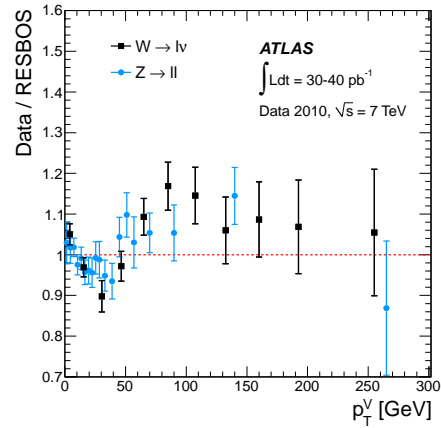} }
\caption{Ratio of the combined electron and muon data to the RESBOS prediction, overlaid with the analogous ratio for the Z transverse momentum distribution measured by ATLAS. }
\label{wzptATLAS}       % Give a unique label
\end{figure}

The Z and W boson production is also studied as a function of the boson $p_T$ in both experiments. Of special interest are the low $p_T$ and high $p_T$ regions of the spectrum. Low $p_T$ region uncertainties are dominated by non-perturbative QCD and the MC tuning. For the Z boson, in CMS~\cite{ZrapidityCMS} the official tune (PYTHIA~\cite{PYTHIA} Z2) gives the best agreement data-MC. In ATLAS~\cite{ZptATLAS} RESBOS~\cite{RESBOS} is in optimal agreement (see Fig.~\ref{zptATLAS}). General disagreement is observed in the high $p_T$ region in both experiments with the theoretical predictions at NNLO. RESBOS in ATLAS and POWHEG~\cite{POWHEG}+PYTHIA Z2 in CMS also show discrepancies in the high $p_T$ region as can be seen in Fig.~\ref{zpthighATLAS}. Something similar happens for the W boson~\cite{WptATLAS}, where a general disagreement in the high $p_T$ region (more than 100 GeV) is observed in ATLAS. MCatNLO~\cite{MCatNLO} and POWHEG+ PYTHIA give a prediction that differs by more than $20\%$ in the high $p_T$ region with respect to data (giving the best prediction in the low domain). Despite the general disagreement shown in these studies, a similar trend is observed in the Z and W boson spectrum (see Fig.~\ref{wzptATLAS}), supporting the universality of strong interactions effects in W and Z production. A better understanding of the $p_T$ spectrum is needed, for example, to achieve the precision required in future W mass measurements. 

\begin{figure}
% Use the relevant command for your figure-insertion program
% to insert the figure file.
% For example, with the option graphics use
\resizebox{0.75\columnwidth}{!}{
  \includegraphics{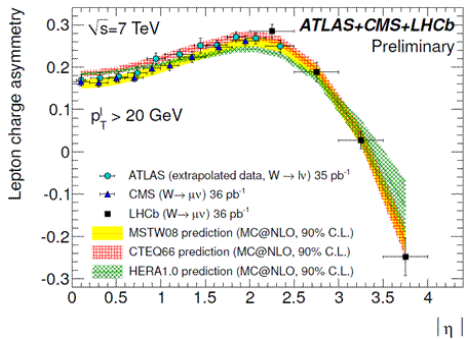} }
\caption{W charge asymmetry as a function of the muon pseudorapidity with data from ATLAS, CMS and LHCb, compared with the MC prediction. }
\label{wasymCMS}       % Give a unique label
\end{figure}

W$^{+}$ and W$^{-}$ are not symmetrically produced in LHC collisions. At $\sqrt{s} =$ 7 TeV the expected ratio of W$^{+}$ over W$^{-}$ is roughly 1.4 and varies with the rapidity of the boson. The accessible variable in LHC experiments to test this asymmetry is the lepton charge asymmetry defined as:
\begin{equation}
A(\eta) = \frac{d \sigma/d\eta(W^{+}\rightarrow l^{+}\nu) - d\sigma/d\eta(W^{-}\rightarrow l^{-}\nu)}{d\sigma/d\eta(W^{+}\rightarrow l^{+}\nu) + d\sigma/d\eta(W^{-}\rightarrow l^{-}\nu)}
\end{equation}
CMS and ATLAS published results for this charge asymmetry with 36 pb$^{-1}$ in the electron and muon channels, (CMS also published results with 234 pb$^{-1}$ in the muon one). They are compared in both experiments with different PDF sets. The best agreement in CMS~\cite{WasymmetryCMS} and ATLAS~\cite{incATLAS} is for the CT10 PDF set. CMS and ATLAS results are also combined with LHCb results, giving information on a wide pseudorapidity range ($|\eta| \leq 4$), see Fig.~\ref{wasymCMS}.

\subsection{Other properties}
\begin{figure}
% Use the relevant command for your figure-insertion program
% to insert the figure file.
% For example, with the option graphics use
\resizebox{0.75\columnwidth}{!}{
  \includegraphics{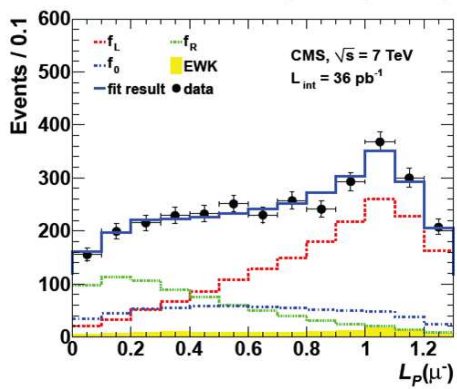} }
\caption{Lepton polarization projection variable for the longitudinal, left and right polarization of the W in the muon channel (CMS).}
\label{lp}       % Give a unique label
\end{figure}

{\bf W polarization.} W bosons at LHC are left polarized at high $p_T$ ($\approx 75\%)$. This is a consequence of the left-handed nature of the electroweak interactions and the prevalence of valence quarks in protons. The lepton projection defined as:
\begin{equation}
L_p = \frac{\vec{p}_T(l) \cdot \vec{p}_T(W)}{|\vec{p}_T(W)|^2}
\end{equation}
is a variable strongly correlated with the W polarization (the correlation increasing with the boson $p_T$). The first observation of the W polarization at the LHC was performed by CMS~\cite{WpolCMS}.To better observe the polarization, only high $p_T$ W bosons are analyzed (larger than 50 GeV). A fit to the lepton projection variable is performed which is sensitive to the three components (left, right and longitudinal) for electrons and muons, see Fig.~\ref{lp}. A left polarization is observed in this study ($\approx 20\%$), reduced with respect to the theoretical one due to kinematic effects and other subprocesses 

%\begin{figure}
% Use the relevant command for your figure-insertion program
% to insert the figure file.
% For example, with the option graphics use
%\resizebox{0.75\columnwidth}{!}{
%  \includegraphics{polCMS.eps} }
%\caption{Left minus right polarization versus the longitudinal polarization of the W as measured in CMS..}
%\label{wpolCMS}       % Give a unique label
%\end{figure}

\vspace{1cm}
\noindent
{\bf Electroweak mixing angle.} This study performed by the CMS collaboration~\cite{weakangCMS} uses 1.1 fb$^{-1}$ of data. In this study, all the electroweak parameters are fixed to their SM values but the effective value of the $\sin \theta_W$ ($\sin \theta_{eff}$) which is left free to vary. A multivariate analysis of the three most sensitive variables to this parameter (boson rapidity, invariant mass and Collins-Soper angle) is used to extract it. The result:
\begin{equation}
\sin \theta_{eff} = 0.2287 \pm 0.002~(stat.) \pm 0.0025~(sys.)
\end{equation}
is in agreement with the SM prediction.

%\begin{equation}
%d \sigma d \eta W^{+} \rightarrow l^{+} \nu
%\end{equation}
% - d\sigma d\eta(W^{-}\ra \cal {l}^{-}\nu)
%{d\sigma/d\eta(W^{+}\ra \cal l^{+}\nu) + d\sigma/d\eta(W^{-}\ra \cal l^{-}\nu)}
%\end{equation}
%
\section{Conclusions}
In this document, the results on W and Z boson properties by both ATLAS and CMS using data from proton-proton collisions at $\sqrt{s}=7$ TeV are presented. In general, they show a good agreement with the Standard Model predictions. A few observed deviations, particulary in the study of the transverse momenta of W and Z bosons, require more dedicated studies.

\end{document}